# Playing Modeling Games in the Science Classroom: The Case for Disciplinary Integration

Pratim Sengupta

University of Calgary

Doug Clark

Vanderbilt University

**Abstract**. We extend the theory of *disciplinary integration* of games for science education beyond the virtual world of games, and identify two key themes of a practice-based theoretical commitment to science learning: (1) materiality in the classroom and (2) iterative design of multiple, complementary symbolic inscriptions (e.g., graphs and agent-based programs). We also identify the affordances of our proposed approach for facilitating student learning and teacher agency.

## Introduction

The Science as Practice perspective (Pickering, 1995; Lehrer & Schauble, 2006) argues that the development of scientific concepts is deeply intertwined with the development of epistemic and representational practices. Clark, Sengupta, Brady, Martinez-Garza, and Killingsworth (2015) proposed *disciplinary integration* as a paradigm for designing and leveraging digital games as a medium to support the development of scientific modeling in K-12 classrooms based on the Science as Practice perspective. In this paper, we identify two key themes of the practice-based theoretical commitment underlying *disciplinary integration*: (1) materiality within the classroom, and (2) iterative design of multiple and complementary symbolic inscriptional systems (e.g., graphs and agent-based programming). We discuss these themes in the context of insights from several research studies that we have conducted with *SURGE NextG* (Clark et al., 2015) and identify the affordances of both these design elements for facilitating student learning and teacher agency.

## The "Science as Practice" Perspective

The Science as Practice (or SaP) perspective (Pickering, 1995; Lehrer & Schauble, 2006; Duschl et al., 2007) argues that the development of scientific concepts is deeply intertwined with the development of epistemic and representational practices (e.g., modeling). Models are inscriptions and fictive representations of real things (e.g., planes, cars, or buildings) or systems (e.g., atomic structure, weather patterns, traffic flow, ecosystems, or social systems), which are simpler than the real objects and systems they represent, and preferentially highlight certain properties of the referent (Rapp & Sengupta, 2012). Modeling is generally recognized as *the* core disciplinary practice in science, and involves the iterative generation and refinement of inscriptions, which in turn serve as, or provide mechanistic explanations of a referent phenomenon (Giere, 1988; Nercessian, 2008; Lehrer & Schauble, 2002).

Science and math education researchers have shown that engaging in modeling and progressively refining one's representation of some aspect of the world (e.g., a model or an inscription) can contribute to a deeper understanding of mathematical and scientific knowledge and practices (Gravemeijer, Cobb, Bowers, & Whitenack, 2000; Hall & Stevens, 1995; Lehrer, 2009; Enyedy, 2005). Lehrer (2009) argues that developing a *good* model involves designing representations that capture essential dynamic features of the relationships they describe, but this also involves learning to foregrounding some elements of the target phenomenon while obscuring or omitting others (Lehrer & Schauble, 2010; Lynch, 1990). Pedagogically, this involves creating iterative opportunities for model evaluation by way of model comparison, i.e., designing curricula that involves the iterative creation of and comparison between different types of inscriptions or models of the same phenomenon (Lehrer & Schauble, 2010; Lesh & Doerr, 2003). This is because models are fundamentally analogies, and therefore, can only be evaluated in light of comparison with competing models (Lehrer & Schauble, 2010).

## Disciplinary Integration and The Design of Digital Games for Learning

Disciplinary integration can be thought of in terms of Allan Collins and colleagues' analyses of "model types" and "modeling strategies" (Collins, White, & Fadel, in preparation), which they have termed "epistemic forms" and "epistemic games" in earlier work (Collins, 2011; Collins & Ferguson, 1993; Morrison & Collins, 1995). Collins and colleagues argue that the professional work of scientists can be understood in terms of epistemic forms (model types that are the target structures guiding scientific inquiry) and epistemic games (modeling strategies that are the sets of rules and strategies for creating, manipulating, and refining those model types).

While Collins and Ferguson did not write with the intention of informing the design of actual digital games (they used the term "game" as a metaphor), disciplinarily-integrated games can be framed around the ideas of Collins and colleagues by structuring digital game play around epistemic games of designing and manipulating formal disciplinary representations (epistemic forms). More specifically, the puzzles and game-play mechanics of disciplinarily integrated games distill epistemic forms (model types) and the epistemic games (modeling strategies) for navigating and manipulating those forms.

The specific emphasis on modeling as game play around disciplinary epistemic forms stands in contrast to engaging in "inquiry" more broadly, as is common in 3D virtual worlds (e.g., *Quest Atlantis, River City,* or *Crystal Island*). A key distinction between these two forms of virtual environments involves the nature and breadth of focus of the inquiry undertaken by students. Virtual inquiry worlds generally engage students in the practices and discourses (Gee, 1990) of a discipline at the level of inquiry writ large. Much of the pedagogical power and engagement of 3D virtual inquiry worlds tends to focus heavily on their impressive affordances for roleplaying, narrative, and identity-building (cf. Gee, 2007; Squire, 2011). While 3D virtual inquiry worlds are compelling and powerful, however, their scope and structure do involve tradeoffs in terms of the individual tasks or puzzles, which are often themselves relatively mundane (e.g., click on a character to be told a piece of evidence, click on a location to get a reading on oxygen levels, or click on a location to bring up another mini-game to collect evidence). Essentially, whereas 3D virtual inquiry worlds tend to cast students as scientists investigating a "mystery" at the level of overarching inquiry, disciplinarily-integrated games do not attempt the depth of immersion, identity-building, and role-playing of virtual inquiry worlds (and do not dispute their importance or value). Instead, disciplinarily-integrated games are designed to engage students more deeply in specific epistemic games of developing and manipulating specific epistemic forms. This focus allows for progressively deepening the puzzle at the heart of the game, and more broadly, all elements of the game, to emphasize the puzzle.

## Examples of Disciplinarily Integrated Games

*SURGE Symbolic* is an example of a disciplinarily integrated game (Figure 1, http://www.surgeuniverse.com). Whereas earlier versions of *SURGE* focused on layering formal representations over informal representations, *SURGE Symbolic* inverts this order, layering informal representations over formal representations while organizing gameplay explicitly around navigating and coordinating across representations. Furthermore, while earlier versions supported *reflection* on the results of game play through formal representations as a means to support strategy-refinement, the formal representations were not the medium through which players *planned, implemented,* and *manipulated* their game strategies. The position graph, for example, can present information about the specific regions of the game-world that will be affected by dangerous electrical storms at given times, as well as about locations where rewards or allies will appear to rendezvous with Surge. As a result of this design approach, the Cartesian space emerges as a set of *scientific instruments* for the player, in the sense of providing access to data about the game world that are not available through other means. At the same time, the Cartesian graphs also play the role of an *instrument panel* or mission planner, offering fine-grained control over the movement of the Surge spacecraft.

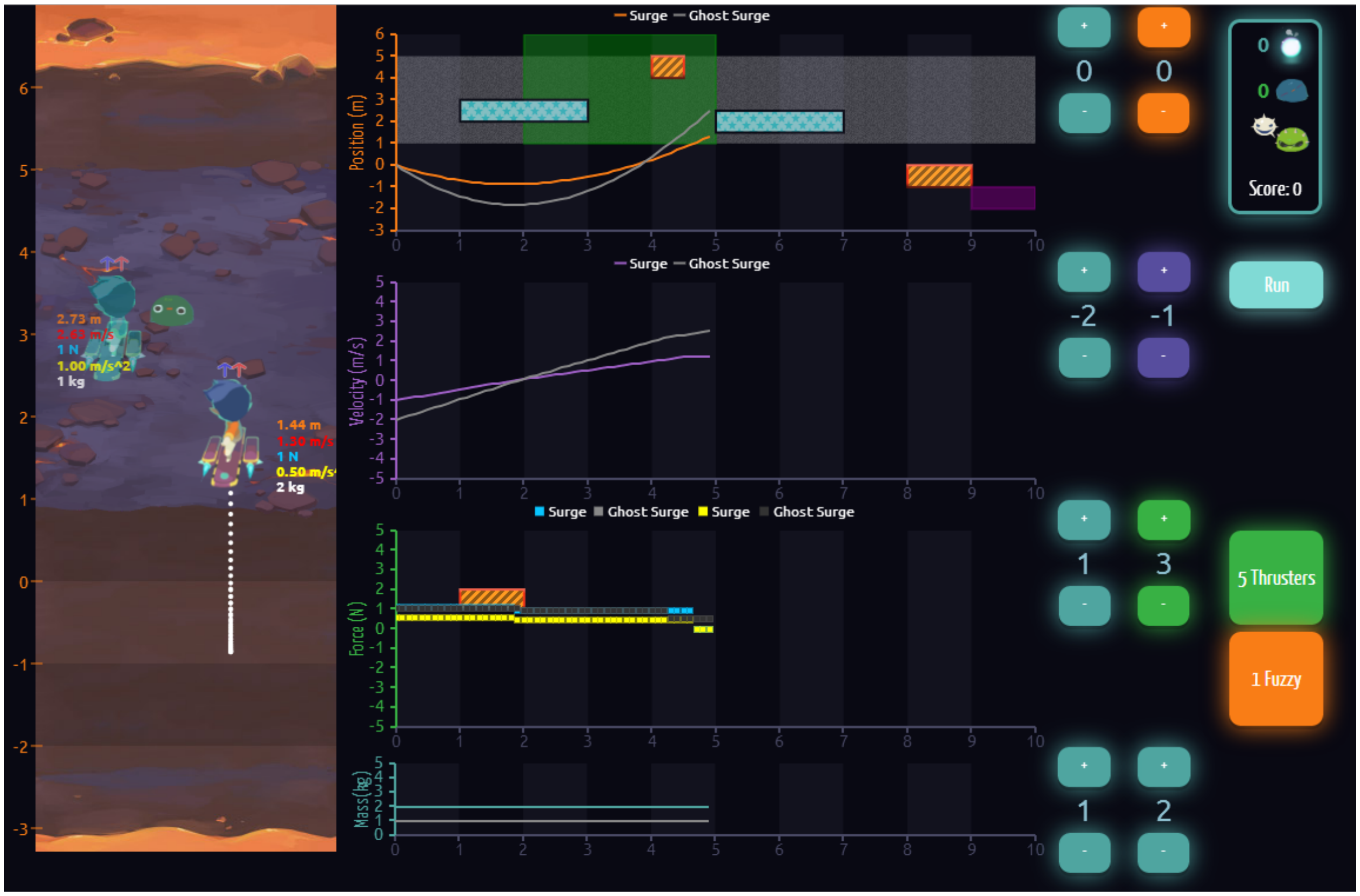

*Figure 1.* SURGE Symbolic

*SURGE Symbolic* is but one possible example for structuring game play around epistemic forms and epistemic games that are authentic (i.e., practices that are at the core of expertise in a target discipline). Framing disciplinary integration in terms of epistemic forms and games allows us to design games in multiple disciplines following the general design approaches proposed here. Clark, Sengupta, and Virk (in press), for example, explore the design of disciplinarily-integrated games across physical science, life science, and social science domains and across multiple model types beyond Cartesian time series analyses (e.g., system dynamics models, situation action models, structured tree models, and constraint system models). Furthermore, disciplinary integration also responds to calls for greater emphasis on problem-solving, 21st century skills, and engaging students in the practices of disciplines to develop deeper understanding. In this paper, our focus is on identifying essential characteristics of a disciplinarily-integrated game that become evident in successful classroom implementations. Of particular interest to us is to identify some characteristics of disciplinarily integrated games, which we believe can enable teachers to seamlessly integrate these games within their curriculum.

## Materiality and Multiple Complementary Inscriptions for Disciplinary Integration

How can disciplinarily-integrated games support the development of modeling as a practice in science classrooms? To address this issue, we have identified two major elements of disciplinary integration that make it possible for students and teachers to engage in scientific modeling activities in the classroom in an authentic manner: (1) material integration of virtual play and (2) the iterative design of and comparison between multiple complementary symbolic inscriptions within and beyond the game's virtual space. In other words, we posit that as students progress through a disciplinarily integrated game, the game should become a sandbox for modeling phenomena in the real world with which they can directly engage in the classroom using material means. Toward this end, while we were developing *SURGE Symbolic*, we began conducting research with *SURGE NextG* (an earlier *SURGE* game) where students sometimes leave the virtual space of the game to conduct modeling-based inquiry in the classroom or in other complementary virtual spaces (e.g., a programming platform designed specifically for the game). These modeling activities are interwoven into the core narrative of the game. When students engage in these activities, they connect actions and interpretations across the multiple virtual and physical spaces, typically by emphasizing the generation of and translation across multiple, complementary inscriptional forms as actions that constitute game-play.

Students in *SURGE NextG* engage in modeling by predicting, generating, navigating between, interpreting, and reflecting upon both spatial records representing position in time (e.g., dot traces) and temporal representations of changes in features of motion over time (e.g., bar graphs). Players navigate their avatar through the play area to rescue game characters and deliver them to safe locations while avoiding obstacles and enemies. In order to move along a particular path, the learner creates a predictive model of the trajectory by placing impulses along the target path (Figure 2). The learner then deploys his or her model by "launching" the level to "run" the plan. Players watch their plans unfold and then revise them accordingly. *SURGE NextG* also includes a graphing environment that enables real-time construction of mathematical representations based on periodic sampling of measures of Surge's motion (see Figure 3).

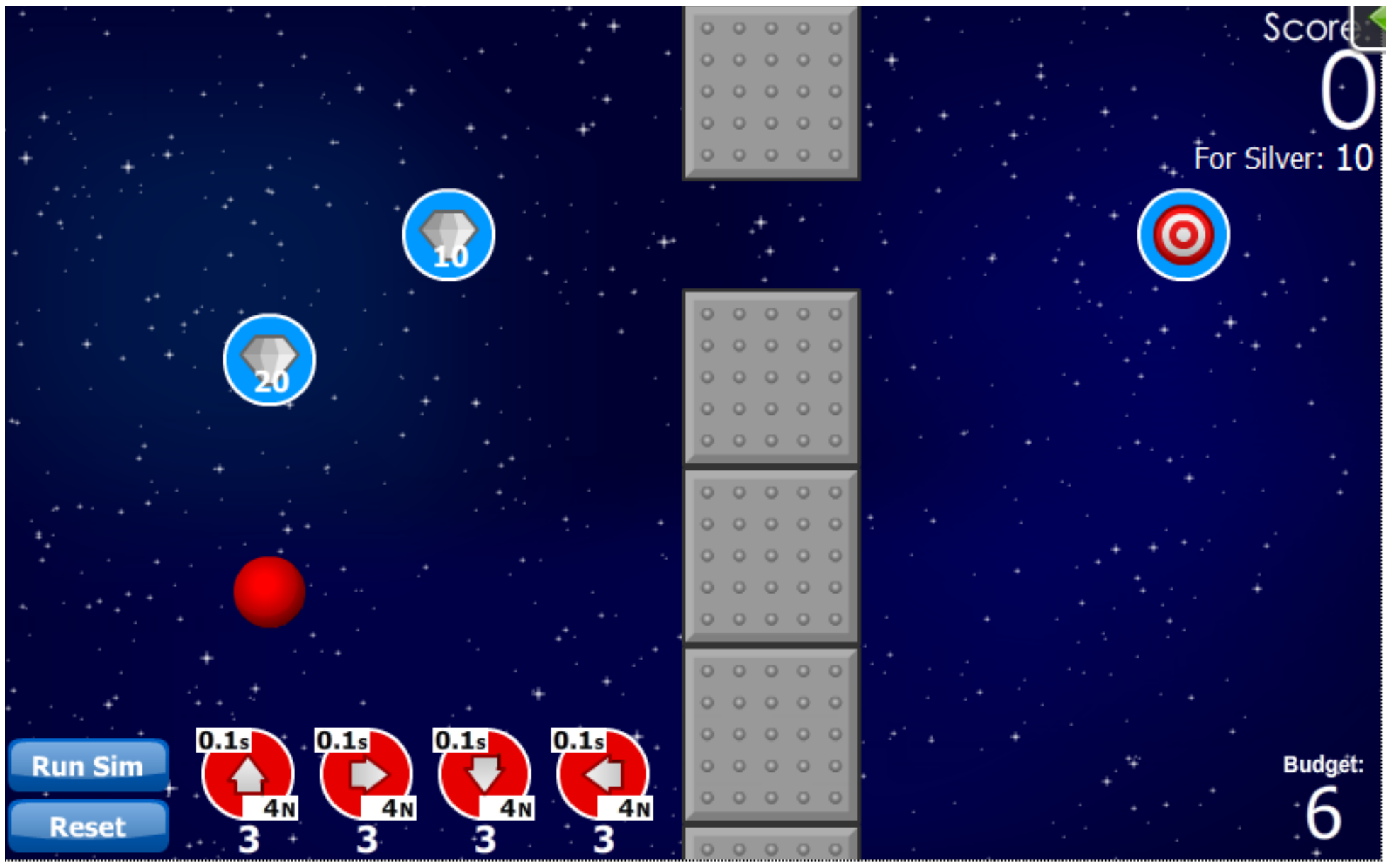

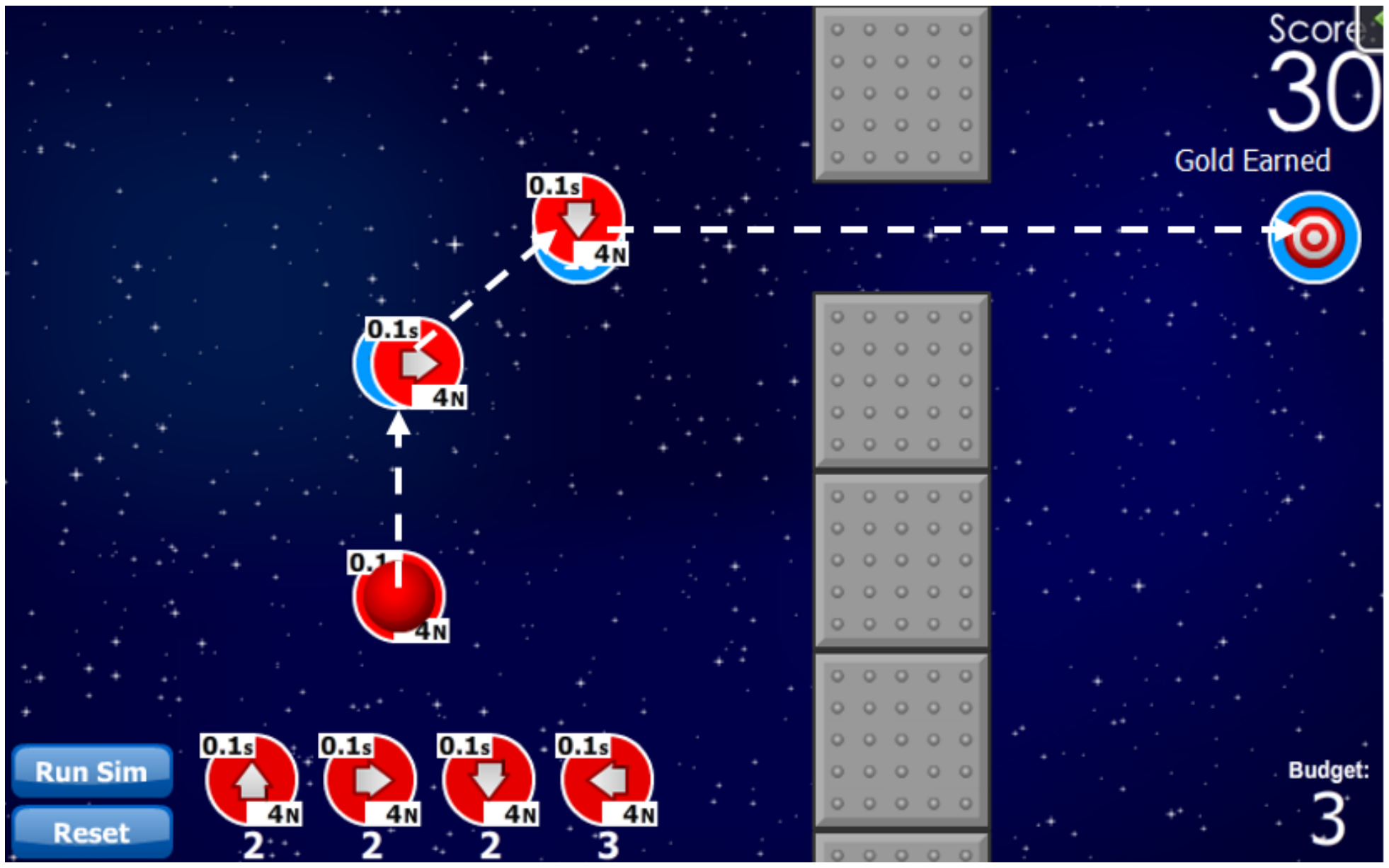

*Figure 2.* Simple *SURGE NextG* level (top) and solution (bottom).

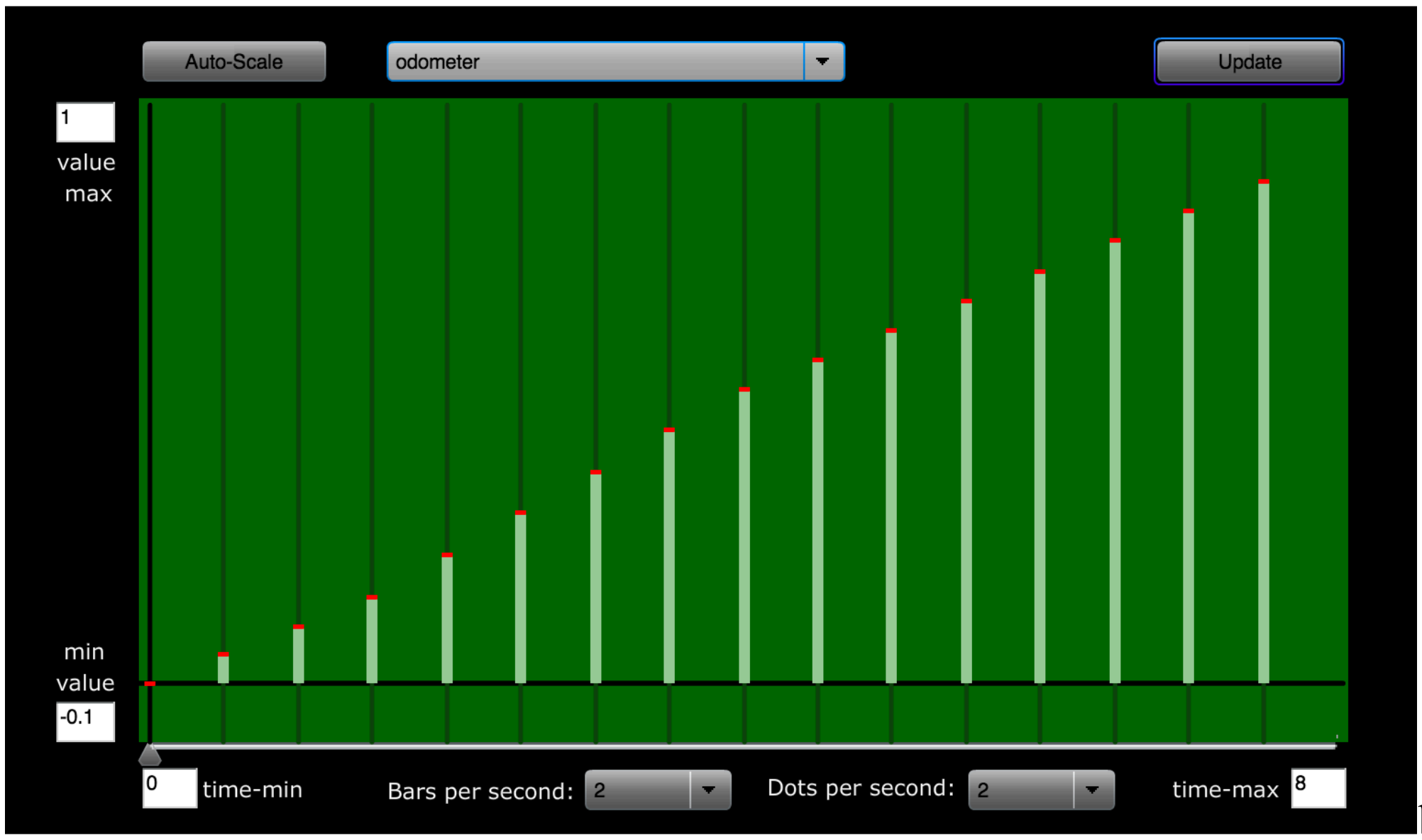

bd

*Figure 3.* The displacement graph that accompanies the one-dimensional motion in *SURGE NextG*.

Our work with *SURGE NextG* has underscored the importance of material integration of games into the classroom, both for students and teachers. For scientists, the process of modeling involves a struggle to get materials and nature to “perform” in the way that they need for their investigations (Pickering, 1995). *Doing* physics involves bringing about alignments of the material world with the world of representation, a process that Pickering (1995) termed the “mangle of practice”. He argued that these alignments are “what sustain specific facts and theories and give them their precise form” (Pickering, 1995, pp 182).

In a recent study (Krinks, Sengupta, & Clark, 2015), for example, students engaged in four class periods of game play within *SURGE NextG* followed by a physical modeling activity that was integrated within the narrative and puzzle at the core of the game. In this modeling activity, students were provided with a marble, track, stopwatch and ruler, and were asked to generate a speed-time graph to represent this motion. Students encountered the “mangle of practice” (Pickering, 1995) as they struggled with the material aspects of inquiry in order to develop conditions for their investigations. That is, students struggled with creating mathematical measures (i.e., units) of speed and acceleration for the rolling ball in terms of its displacement in equal intervals of time. This act of measurement involved bringing into alignment the following: reasoning about what materials to use for measurement; material properties, such as steepness of the ramp; how to use the materials (e.g., how often to take measurements of time); and their theories and assumptions about what would constitute a good measurement of speed. This “mangle” was then extended to the virtual world of the game, when they returned to the game environment, and were asked to help Surge navigate through a nebula that was interfering with communication. In order to successfully guide Surge’s ship and follow certain parameters, they received a “clue” from the captain to help them with their navigation task—a video of a ball rolling down a ramp and then up another ramp until it stopped at its maximum height (Figure 4). This video clue was designed specifically to be similar to their work from the previous day (the marble rolling down the ramp). They were then asked to design a navigation plan for Surge within the virtual world so that the pattern of changes in Surge's speed matched the changes in speed of the ball rolling down and up the ramp (Figure 5). A successful completion of this level involved reasoning explicitly about how speed was being measured in SURGE Next (in the form of dot traces and graphs), so that they could place impulses and forces appropriately along the desired trajectory, and complete their mission of getting Surge's ship safely through the nebula.

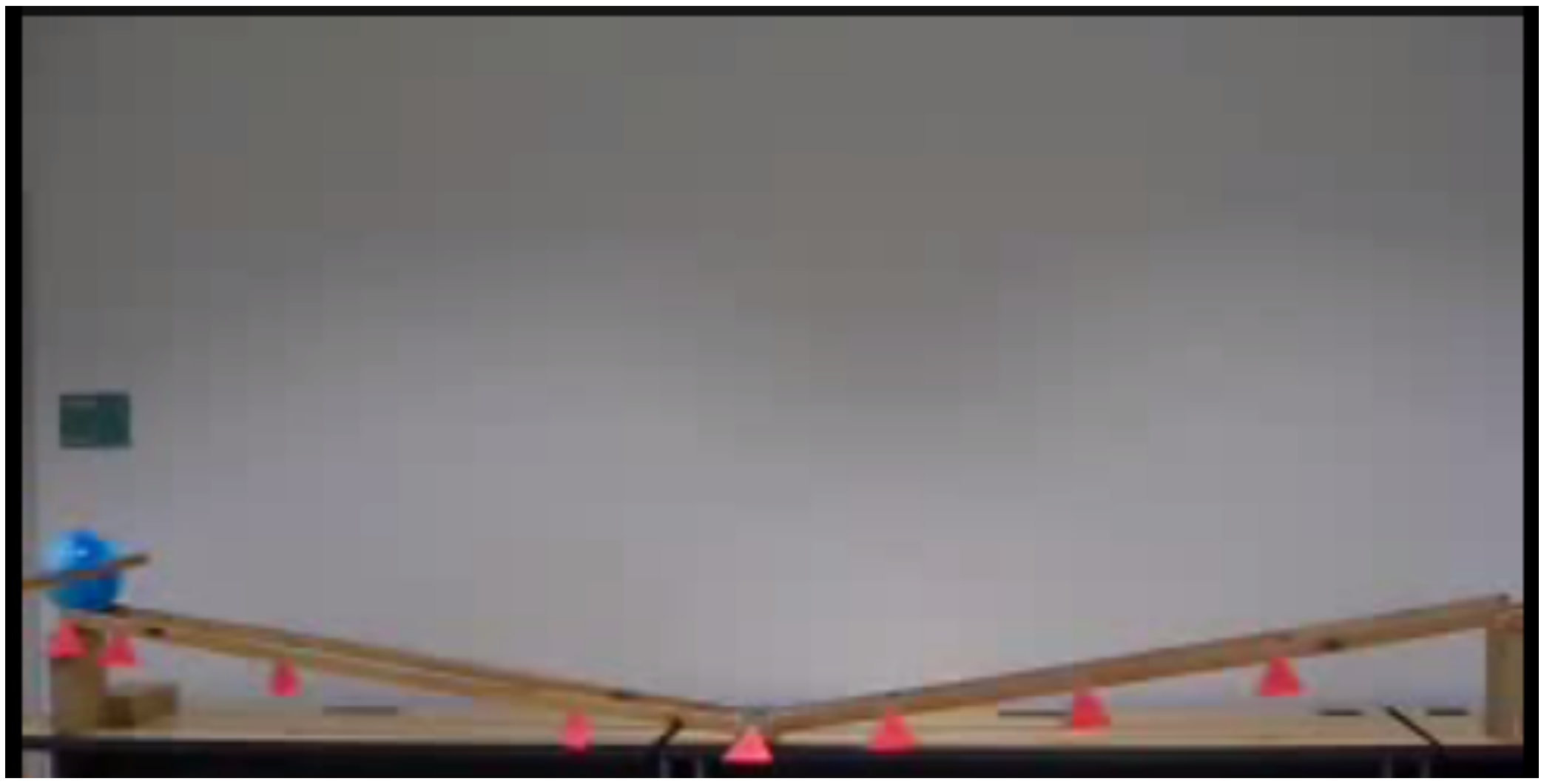

*Figure 4.* Screenshot from a real-world video "Clue" embedded in SURGE NextG

Another study with *SURGE NextG* highlighted the importance of complementary inscriptions (Sengupta, Clark, Krinks, Killingsworth, & Brady, 2014). Each class section was assigned to one of two graphing conditions (grapher vs. no grapher). Students in the grapher sections showed significantly greater pre-post test gains on key Newtonian relationships pertaining to Newton's second law of motion. Furthermore, the grapher students who framed their plan in a variable-focused way (e.g., by explicitly reasoning about the *meaning* of the relationships between distance, speed and acceleration evident in the graph) rather than superficially (e.g., by focusing only on the *shape* of the graph) showed the highest gains of all. As students progressed through the levels of *SURGE NextG*, the game became a modeling tool for iteratively developing formal representations (graphs of motion) and connecting these representations to situations of motion in the real world. In the earlier levels, the game play was designed to familiarize students with the core computational representation of motion – dot traces – by engaging them in designing predictive trajectories of Surge. In later levels, students iteratively designed inscriptions in *SURGE NextG* in order to further mathematize the dot-trace representations, and used these two representations to model an example of constant acceleration in the real world. In doing so, students developed explanations (models) of motion, made predictions, and deployed their models to verify and further improve on them – and in the process, developed a deep understanding of conceptual relationships in kinematics. This relationship between meaningful symbolization and the development of conceptual understanding is central to disciplinary integration.

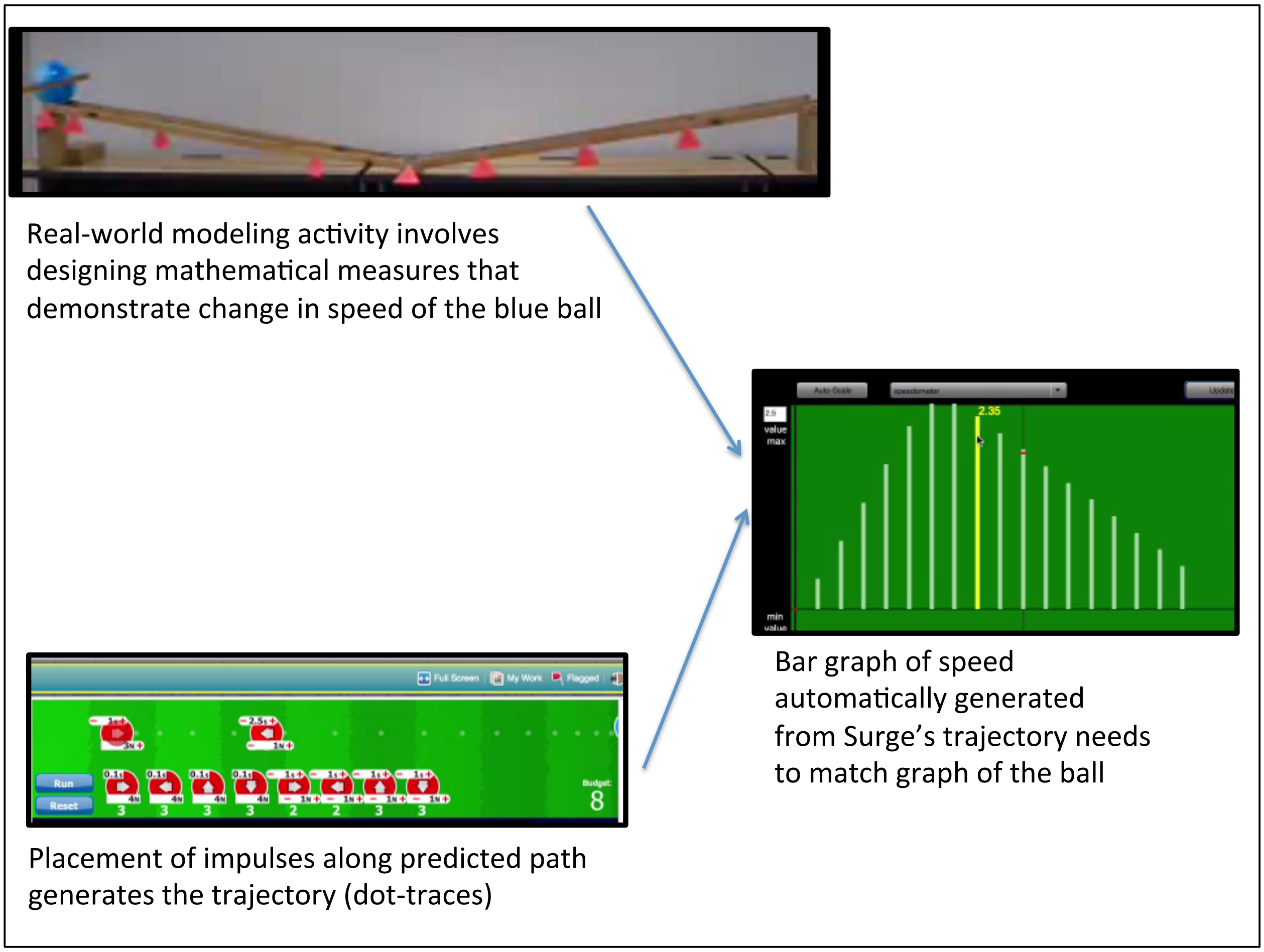

*Figure 5.* Integration of real-world modeling with SURGE NextG

More recently, we have investigated the further addition of another complementary inscriptional tool to *SURGE NextG*: agent-based programming. Students in these studies used *ViMAP*, a visual programming environment that provides students with domain-specific and domain-general programing primitives along with flexible and powerful graphing functionality (Sengupta et al., 2015). Our *ViMAP* research showed that even elementary grade students can begin to understand motion as a process of continuous change by (a) modeling the behavior of the agent that produces a dot-trace representation using agent-based programming commands and (b) graphing the motion of the agent in a manner that makes explicit the pattern of change over time (Farris & Sengupta, 2014; Sengupta, Farris & Wright, 2012).

We designed a version of the *ViMAP* programming language specifically to visually represent Surge as the computational agent. One might argue that layering programming on top of game play might be challenging for learners. We found that when interwoven within the core puzzle of the game, however, agent-based programming can serve as a reflexive activity that can deepen students' conceptual understanding of the underlying physics. In terms of the game narrative, *ViMAP* served as a "garage" for Surge's ship. Students had to generate a graph of varying acceleration to fix the graphing engine for Surge and, in the following activity, had to reproduce the graph in *SURGE NextG* (i.e., by placing impulses on the desired trajectory) to demonstrate the efficacy of the new engine to the game characters (Figure 6). In-depth microgenetic case studies demonstrated that as learners engaged with generating and translating across complementary inscriptional forms, they were able to develop progressively refined explanations about the semantic relationship between computational actions and physics concepts (e.g., the step-size of the ViMAP agent represents speed of Surge), and the underlying mathematical relationships between the physics concepts (e.g., the amount of change in step-size represents acceleration).

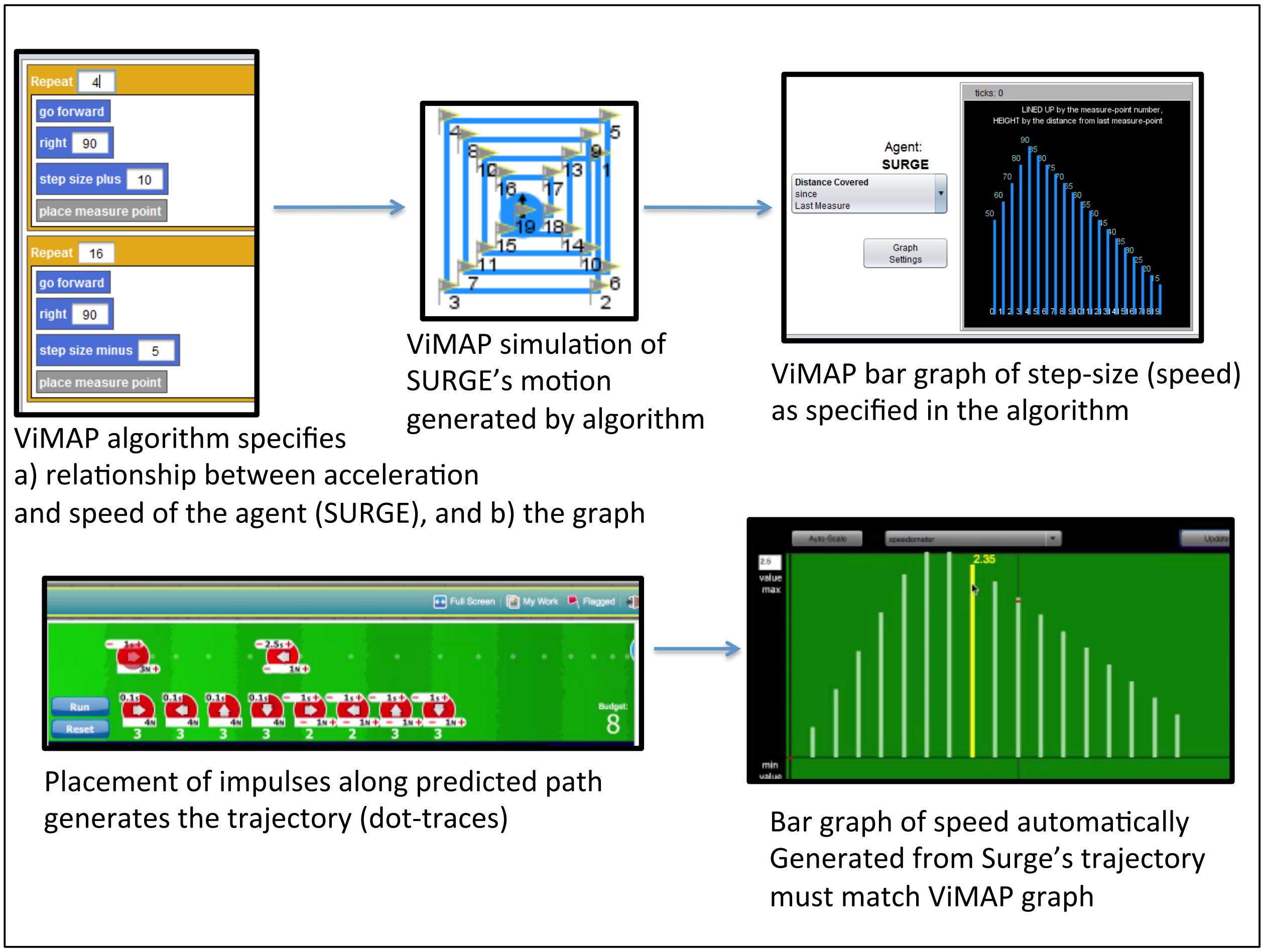

*Figure 6.* Integration of agent-based programming with SURGE NextG

## Final Thoughts: Implications for Game Designers and Teachers

Disciplinary integration of games for science focuses on a key epistemic and representational practice: modeling. While much has been written about the pedagogical virtue of digital games, wide-spread successful adoption and implementation in classrooms remains elusive. We believe that integration of games in science classrooms needs to simultaneously account for the role of the teacher, and integration of the activities and conceptual developments that constitute gameplay within the larger ecosystem of the curriculum. To this end, we have argued that the emphasis on modeling as gameplay can be operationalized as follows:  a) by integrating the virtual modeling activities within the material world of the classroom and b) by emphasizing the design of, and comparison between, multiple complementary inscriptional systems as key elements of game play. Viewed in this light, disciplinarily-integrated games become sandboxes for modeling scientific phenomena across contexts, inscriptions, and practices.

The integration of games within the classroom fundamentally relies on the teacher and the students being able to meaningfully adopt, understand, value, and implement the game as a both a context and a tool for scientific modeling. This in turn involves supporting the students in successfully bridging ideas across contexts and representations. Therefore, we strongly believe that game designers *must* pay attention to the issues of designing complementary mathematical inscriptions, and creating opportunities to connect virtual game play with modeling activities in the real world. This is important both from the perspective of students and teachers. When students create and translate across complementary inscriptions (such as dot-traces, graphs and agent-based algorithms) of the same phenomenon as part of the core puzzle at the heart of the game, they deepen their conceptual understanding of science, and begin to see the game as a *model* of the physical world.

From the perspective of teachers, both of the characteristics of disciplinary integration – materiality and multiple complementary representations - played important roles. Along the first dimension, as several of the teachers in our studies reported to us, connecting the physical and virtual worlds provided them with productive opportunities to appropriate the game as part of their broader curricular goals. This enabled them to integrate games within their regular and longer-term classroom practice. In the studies described earlier, the teachers solicited students' ideas about how to measure speed, introduced formal language to re-describe students' intuitive ideas, and most importantly, led class discussions to develop a normative way of measuring speed. The teachers found these activities familiar because of their familiarity with the physical objects used for the modeling activity (ramps, balls, tapes), the phenomenon (rolling down), the underlying mathematical relationships (constant acceleration as continuous change over time), and mathematical representations (graphs). Furthermore, we also found for all teachers that generating and interpreting graphs is a key focus in their regular math and science curriculum throughout the upper elementary and middle grades, and that representing motion as a process of change over time can help teachers and students work toward this goal.

In order to support disciplinary integration, our research thus demonstrates that designers of educational games should focus on core epistemic practices, integrate these practices within game-play through central complementary inscriptional systems, and create opportunities to integrate the material world and virtual game through these complementary inscriptional systems and practices so that teachers can integrate the game within their curriculum.

**Bios**

Pratim Sengupta, assistant professor of the learning sciences at University of Calgary, designs and develops agent-based computational technologies for long-term, ecologically valid classroom integration, empirically grounded models of short-term conceptual change and long-term development of learners, and multi-agent models of complex social and natural phenomena. While at Vanderbilt, he received an NSF CAREER award on integrating programming in K12 science classrooms.

Pratim@vimapk12.org

Doug Clark, professor of the learning sciences and science education at Vanderbilt, investigates the learning processes through which people come to understand core science concepts in the context of digital learning environments and games. This work focuses primarily on conceptual change, inquiry, modeling, explanation, collaboration, and argumentation. He is principal investigator on NSF and DOE grants on games.

doug.clark@vanderbilt.edu